\documentclass[10pt,a4paper]{article}
\usepackage[utf8]{inputenc}
\usepackage{amsmath, amsthm, amssymb}
\usepackage{enumerate}
\newtheorem{thm}{Theorem}[section]

\newtheorem{tvrzx}[thm]{Proposition}

\newtheorem{lemmax}[thm]{Lemma}

\newtheorem{theoremx}[thm]{Theorem}

\theoremstyle{definition}
\newtheorem{definicex}[thm]{Definition}

\theoremstyle{remark}
\newtheorem{remx}[thm]{Remark}

\theoremstyle{definition}
\newtheorem{examplex}[thm]{Example}

\def\R{\mathbb{R}}

\def\<{\langle}
\def\>{\rangle}
\def\~{\widetilde}
\def\^{\wedge}

\def\io{\mathit{i}}

\newcommand{\bm}[4]{ \left( \begin{array}{cc} #1 & #2 \\ #3 & #4 \end{array} \right) }
\newcommand{\bv}[2]{ \left( \begin{array}{c} #1 \\ #2 \end{array} \right) }
\newcommand{\bs}[1]{ \boldsymbol{#1} }

\begin{document}
\begin{flushright}
\today
\end{flushright}
\vspace{0.7cm}
\begin{center}

\baselineskip=13pt {\Large \bf{On the Generalized Geometry Origin of Noncommutative Gauge Theory}\\}
 \vskip0.5cm
 {\it dedicated to Bruno Zumino on the occasion of his 90th birthday}  
 \vskip1.3cm
 Branislav Jur\v co$^1$, Peter Schupp$^2$, Jan Vysoký$^{2,3}$\\
 \vskip0.6cm
$^{1}$\textit{Mathematical Institute, Faculty of Mathematics and Physics,
Charles University\\ Prague 186 75, Czech Republic;\footnote{permanent address}\\
CERN, Theory Division, CH-1211 Geneva 23, Switzerland, jurco@karlin.mff.cuni.cz}\\
\vskip0.3cm
$^{2}$\textit{Jacobs University Bremen\\ 28759 Bremen, Germany, p.schupp@jacobs-university.de}\\
\vskip0.3cm
$^{3}$\textit{Czech Technical University in Prague\\ Faculty of Nuclear Sciences
and Physical Engineering\\ Prague 115 19, Czech Republic, vysokjan@fjfi.cvut.cz}\\
\vskip0.5cm
\end{center}
\vspace{0.4cm}

\begin{abstract}
We discuss noncommutative gauge theory from the generalized geometry
point of view. We argue that the equivalence between the commutative
and semiclassically noncommutative DBI actions is naturally encoded
in the generalized geometry of D-branes.
\end{abstract}

{\textit{Keywords:}} Bosonic Strings, Poisson Sigma Models,
D-Branes, Courant-Dorfman Brackets, Dirac-Born-Infeld Action,
Seiberg-Witten map, Non-commutative Gauge Theory.

\section{Introduction}
Generalized geometry \cite{Hitchin:2004ut,Gualtieri:2003dx} recently
appeared to be a powerful mathematical tool for the description of
various aspects of string and field theories. Here we mention only
few instances of its relevance that are more or less directly related to the present
paper. Topological and non-topological Poisson sigma models are
known to be intimately related to a lot of interesting differential,
in particular generalized, geometry. For instance, the topological
Poisson sigma models are of interest for the integration of Poisson
manifolds (and Lie algebroids) \cite{2004LMaPh..67...33C} and are at the heart of deformation quantization
\cite{Cattaneo:1999fm}. Field equations of (topological) Poisson sigma models can
be interpreted as Lie algebroid morphisms \cite{liemorphisms} and as
such can further be generalized in terms of generalized (complex)
geometry \cite{kotov}, \cite{kotovstrobl}. Poisson sigma models can be
twisted by a 3-form $H$-field \cite{Klimcik:2001vg} and also generalized to Dirac
sigma models \cite{kotovstrobl}, where the graph defined by the corresponding
(possibly twisted) Poisson structure is replaced by a more general
Dirac structure. In turn, at least in some instances, D-branes can be related to Dirac structures \cite{SeveraLetter2},
\cite{Asakawa:2012px}, or coisotropic submanifolds \cite{2004LMaPh..69..157C}. In \cite{alekseevstrobl}, it has been
observed that the current algebra of sigma models naturally involves
structures of generalized geometry, such as the Dorfman bracket and
Dirac structures. This was further developed in \cite{zabzine} and
\cite{zabzinbonelli}. In \cite{halmagyi}, it was observed that in
the first order (nontopological) Poisson sigma model characterized
by a $2$-form $B$ and a bivector $\theta$, a more general form of world-sheet currents appears. Their algebra has been shown to close under
a more general bracket, the so called Roytenberg bracket \cite{roytenberg_quasi}. In
\cite{halmagyi2}, it has been shown that the structure constants of
the Roytenberg bracket appear if one lifts the topological part of
first order Poisson sigma characterized by a $2$-form $B$ and a
bivector $\theta$ to a three-dimensional WZW term. It this respect,
generalized geometry is relevant for discussions of
non-geometric backgrounds.

Noncommutativity of open strings, more precisely of their endpoints,
in the presence of a $B$-field was recognized in \cite{Chu:1998qz},
\cite{Schomerus:1999ug} and \cite{Ardalan:1998ce}. A thorough
discussion of noncommutativity in string theory followed in the
famous article of Seiberg and Witten \cite{Seiberg:1999vs}, where,
among other things, also the equivalence of commutative and
noncommutative gauge theories was discussed via a field redefinition
known under the name Seiberg-Witten map. In particular, it was
argued that the higher derivative terms in the noncommutative
version of the Dirac-Born-Infeld (DBI) action can be viewed as
corrections to the usual DBI action, the effective D-brane action.
For reviews on noncommutativity in string theory we refer, e.g., to \cite{Douglas:2001ba},
\cite{Szabo:2001kg}. Let us also note that the (semiclassical) noncommutativity
of D-branes can be seen as the (semiclassical) noncommutativity of
the string endpoints in the open topological Poisson sigma model
\cite{2004LMaPh..67...33C}, which fits naturally to their role in both the
integration as well as deformation quantization of Poisson
structures.

The purpose of the present paper is to unravel the generalized
geometry origin of noncommutative gauge theory. We will mainly
focus on the equivalence between the commutative and semiclassically
noncommutative DBI actions (and closely related issues) and argue
that the necessity of such an equivalence can be seen and naturally
interpreted within generalized geometry. In the discussion,
non-topological Poisson sigma models play a role. Roughly
speaking, we intend to convince the reader that the
equivalence of  commutative and semiclassically noncommutative DBI
actions is encoded in two different ways of expressing a
generalized metric on a D-brane.

Before going into a more detailed description of the individual
sections, let us note that almost everything in this paper is
presented in a form suitable for a direct generalization to Nambu-Poisson
structures and M-theory membranes, cf. \cite{Jurco:2012yv}, \cite{Schupp:2012nq}.
We will discuss this in detail in a forthcoming paper.

The paper is organized as follows:

In the second section, we review basic definitions of generalized
geometry. We emphasize the behavior of a generalized metric under
orthogonal transformations of $TM \oplus T^{\ast}M$. This allows us
to recover the formulas relating, via a bivector
$\theta$, the closed background fields $g$, $B$ and the open string
backgrounds $G$ and $\Phi$. It comes as a relation between two generalized
metrics, which are connected by the action of a certain orthogonal
transformation induced by the bivector $\theta$. Finally, we recall
the definition of the Dorfman bracket, Dirac structures and their
relation to D-branes. In the latter we follow the proposal of \cite{Asakawa:2012px}, where D-branes correspond to leaves of foliations defined by Dirac structures.

In the third section, we observe that adding the gauge field $F$ on
D-brane volume corresponds to an action of an orthogonal
transformation on the natural generalized metric on the D-brane, the
pullback of the generalized space-time metric defined by the closed
backgrounds $g$ and $B$. The natural question is whether the so
obtained generalized metric can again be rewritten in the open
string variables (with some gauge field $F'$ and a possibly
modified bivector $\theta'$). The positive answer is given by two
different factorizations of an orthogonal transformations defined by
a bivector and a 2-form, in our case $\theta$ and $F$. As a
consequence, we find a generalization of open-closed relations of
Seiberg and Witten, which includes the field strengths $F$ and $F'$,
the latter one closely related to the nocommutative gauge field
strength. This equality, crucial for our discussion of DBI actions,
also hints towards the appearance of the semiclassical
Seiberg-Witten map, once one recalls its interpretation as the local
coordinate change between the two (Poisson) bivectors $\theta$ and
$\theta'$.

In the fourth section, we use the above mentioned relation between open and closed
variables (including gauge fields) to show that non-topological
Poisson-sigma model, its
Hamiltonian and the corresponding Polyakov action  are manifestly invariant under the open-closed field
redefinitions as they geometrically correspond to the same
generalized metric.

In the fifth section, we briefly recall the interpretation of the
semiclassical Seiberg-Witten map as a local diffeomorphism on the
D-brane world volume relating the noncommutativity parameters (Poisson
bivectors) $\theta$ and $\theta'$. This interpretation is the most
relevant one for our discussion in the final section. When considering D-branes which are symplectic leaves of $\theta$, the Seiberg-Witten map is naturally interpreted in terms of the corresponding Dirac structure.

In the final section, we discuss the equivalence of commutative and
semiclassically noncommutative DBI action of a D-brane. We show that this
equivalence is a direct consequence of the (gauge field dependent)
open-closed relations combined with a Seiberg-Witten map. The
discussion here is not completely new. However, what we believe is
new and interesting is the clear generalized geometry origin of its
main ingredients as developed in previous sections. Everything works very naturally for a D-brane which is a symplectic leaf of the Poisson structure, describing the noncommutativity.

We believe that analogous results hold also for more general $D$-branes, i.e. those which are related to more general Dirac structures than the ones defined by graphs  of Poisson tensors. For such D-branes, Dirac sigma models of \cite{kotovstrobl} should replace the Poisson sigma models.

\section{Generalized geometry}
\subsection{Fiberwise metric, generalized metric}
In this section we recall some basic facts regarding generalized
geometry, see, e.g., \cite{Gualtieri:2003dx}, \cite{Bouwknegt:2010zz}. Although most of the involved objects can be defined in a
more general framework, we focus on a particular choice of vector
bundle. Namely, let $M$ be a smooth manifold and $E = TM \oplus
T^{\ast}M$. A fiberwise metric $(\cdot,\cdot)$ on $E$ is a
$C^{\infty}(M)$-bilinear map $(\cdot,\cdot): \Gamma(E) \times
\Gamma(E) \rightarrow C^{\infty}(M)$, such that for each $p \in M$,
$(\cdot,\cdot)_{p} : E_{p} \times E_{p} \rightarrow \R$ is a
symmetric non-degenerate bilinear form. There exists a canonical
fiberwise metric $\<\cdot,\cdot\>$ on $E$, defined as
\begin{equation}
\label{def_canmetric} \< V + \xi, W + \eta \> = \io_{V}(\eta) +
\io_{W}(\xi), \end{equation} for every $(V + \xi), (W + \eta) \in
\Gamma(E)$. This fiberwise metric has signature $(n,n)$, where $n$
is a dimension of $M$. Hence, we denote by $O(n,n)$ the set of
vector bundle automorphisms preserving this fiberwise metric. That
is
\begin{equation} O(n,n) = \{ O \in \Gamma(Aut(E)) \ | \ (\forall e_{1},e_{2}
\in \Gamma(E)) \ (\<Oe_{1},Oe_{2}\> = \<e_{1},e_{2}\>) \}. \end{equation}
There are three important examples of $O(n,n)$ transformations,
which we will use in the sequel. Let $B \in \Omega^{2}(M)$ be a
$2$-form on $M$. In this paper we will always denote the induced
vector bundle morphism from $TM$ to $T^{\ast}M$ by the same letter,
i.e., we define
\begin{equation} B(V) = -\io_{V}B = B(\cdot,V), \end{equation}
for all $V \in \mathfrak{X}(M)$. Correspondingly, the map $e^{B}$ is
given as
\begin{equation} e^{B}(V + \xi) = V + \xi + B(V).
\end{equation} In the block matrix form
\begin{equation} e^{B}\bv{V}{\xi} = \bm{1}{0}{B}{1} \bv{V}{\xi}, \end{equation}
for all $(V+\xi) \in \Gamma(E)$. Similarly, let $\theta \in \Lambda^{2}
\mathfrak{X}(M)$ be a bivector. The induced vector bundle morphism
is again denoted by the same letter, that is
\begin{equation} \theta(\xi) := -\io_{\xi} \theta = \theta(\cdot,\xi), \end{equation}
for all $\xi \in \Omega^{1}(M)$. Correspondingly, we have $e^{\theta}$
\begin{equation} e^{\theta}(V + \xi) = V + \xi + \theta(\xi). \end{equation}
In the block matrix form
\begin{equation} e^{\theta}\bv{V}{\xi} = \bm{1}{\theta}{0}{1} \bv{V}{\xi}, \end{equation}
for all $(V+\xi) \in \Gamma(E)$. Finally, let $N: TM \rightarrow TM$ be any invertible smooth vector bundle morphism over identity. We define the map $O_{N}$ as
\begin{equation} O_{N}(V+\xi) := N(V) + N^{-T}(\xi), \end{equation}
where $N^{-T}: T^{\ast}M \rightarrow T^{\ast}M$ denotes the map transpose to $N^{-1}$. In the block matrix form
\begin{equation} O_{N}\bv{V}{\xi} = \bm{N}{0}{0}{N^{-T}} \bv{V}{\xi}. \end{equation}
Any $O(n,n)$ transformation with the invertible upper-left block can
be uniquely decomposed as a product of the form \begin{equation}
\label{decomp} e^{-B}O_N e^{-\theta}.
\end{equation}
More explicitly, for $\bm{A_{11}}{A_{12}}{A_{21}}{A_{22}}$ in
$O(n,n)$, i.e., $A_{21}^TA_{11} + A_{11}^TA_{21}=0$, $A_{12}^TA_{22}
+ A_{22}^TA_{12} =0$ and $A_{21}^TA_{12} + A_{11}^TA_{22} =1$, we
find $N = A_{11}$, $\theta = - A_{11}^{-1}A_{12}$ and
$B=-A_{21}A_{11}^{-1}$.

Let now $\tau: \Gamma(E) \rightarrow \Gamma(E)$ be a
$C^{\infty}(M)$-linear map of sections, such that $\tau^{2} = 1$.
For $e_{1},e_{2} \in \Gamma(E)$, we put
\begin{equation} (e_{1},e_{2})_{\tau} := \<\tau(e_{1}), e_{2} \>.
\end{equation}
If such $(.,.)_{\tau}$ defines a  positive definite fiberwise
metric, we refer to it as a generalized metric on $E$. From now on,
we will always assume that this is the case. Since
$(\cdot,\cdot)_{\tau}$ is symmetric, $\tau$ is a symmetric map, that
is,
\begin{equation} \<\tau(e_{1}),e_{2}\> = \<e_{1},\tau(e_{2})\>, \end{equation}
for all $e_{1},e_{2} \in \Gamma(E)$. Also, because $\tau^{2} = 1$,
it is orthogonal and thus $\tau \in O(n,n)$. Moreover, from
$\tau^{2} = 1$, we get two eigenbundles $V_{+}$ and $V_{-}$,
corresponding to $+1$ and $-1$ eigenvalues of $\tau$, respectively.
Using the fact that $(\cdot,\cdot)_{\tau}$ is positive definite, we
get that $\<\cdot,\cdot\>$ is positive definite on $\Gamma(V_{+})$
and negative definite on $\Gamma(V_{-})$. Finally, we can observe
that $V_{+}^{\perp} = V_{-}$ with respect to $\<\cdot,\cdot\>$ and
vice versa, and using the knowledge of the signature of
$\<\cdot,\cdot\>$, we get the direct sum decomposition
\begin{equation} E = V_{+} \oplus V_{-}. \end{equation}
Conversely, for any subbundle $V$ of $E$ of rank $n$, on which
$\<\cdot,\cdot\>$ is positive definite, set $\tau|_V := +1$ and
$\tau|_{V^{\perp}}= -1$ to get a generalized metric on $E$.

From positive definiteness on $V_{+}$, we have $V_{+} \cap TM = 0$
and $V_{+} \cap T^{\ast}M = 0$, and the same for $V_{-}$. This means
that $V_{+}$ and $V_{-}$ can be viewed as graphs of invertible
smooth vector bundle morphisms:
\begin{equation} V_{+} = \{ V + A(V) \ | \ V \in TM \} \equiv \{A^{-1}(\xi) + \xi \ | \ \xi \in T^{\ast}M \}, \end{equation}
\begin{equation} V_{-} = \{ V + A'(V) \ | \ V \in TM \} \equiv \{A'^{-1}(\xi) + \xi \ | \ \xi \in T^{\ast}M \}, \end{equation}
where $A,A': TM \rightarrow T^{\ast}M$, respectively. We can view
$A$ as covariant 2-tensor field on $M$, and write uniquely $A = g +
B$, where $g$ is a symmetric part of $A$ and $B$ a skew-symmetric
part of $A$. From the positive definiteness of $V_{+}$ we get that
$g$ is a Riemannian metric on $M$, whereas $B$ can be an arbitrary
$2$-form on $M$. Using the orthogonality of $V_{+}$ and $V_{-}$, we
see that $A' = -g + B$. From this equivalent formulation, i.e. using $g$
and $B$, we can uniquely reconstruct $\tau$. This will give
\begin{equation} \tau(V + \xi) = (g  - Bg^{-1}B)(V) - g^{-1}B(V) + Bg^{-1}(\xi) + g^{-1}(\xi), \end{equation}
for all $(V + \xi) \in \Gamma(E)$. In the block matrix form,
\begin{equation} \tau \bv{V}{\xi} =  \bm{-g^{-1}B}{g^{-1}}{g - Bg^{-1}B}{Bg^{-1}} \bv{V}{\xi}. \end{equation}
The corresponding fiberwise metric $(\cdot,\cdot)_{\tau}$ can then
be written in the block matrix form
\begin{equation} \label{gm_matrix} (V+\xi,W+\eta)_{\tau} = \bv{V}{\xi}^{T} \bm{g - Bg^{-1}B}{Bg^{-1}}{-g^{-1}B}{g^{-1}} \bv{W}{\eta}. \end{equation}

The important observation is that the block matrix in formula
(\ref{gm_matrix}) can be written as a product of simpler matrices.
Namely,
\begin{equation} \bm{g - Bg^{-1}B}{Bg^{-1}}{-g^{-1}B}{g^{-1}} = \bm{1}{B}{0}{1} \bm{g}{0}{0}{g^{-1}} \bm{1}{0}{-B}{1}. \end{equation}

Note the important fact that the $2$-form $B$ \emph{does not} have to be closed, and this will remain true throughout the whole paper. Nevertheless, we assume that $B$ is globally defined, i.e. $H=dB$ globally.\footnote{More precisely, we assume that the corresponding integral cohomology class $[H]$ is trivial.} We thus consider only the models with trivial $H$-flux. The case of the non-trivial $H$-flux will be discussed elsewhere.
 
There exists a natural action of the group $O(n,n)$ on the space of
generalized metrics. For each $O \in O(n,n)$ and given $\tau$ define
$\tau' = O^{-1} \tau O$. Clearly $\tau'^{2} = 1$ and
\[ \< \tau'(e_{1}), e_{2} \>=  \<
\tau(O(e_{1})), O(e_{2}) \> = (O(e_{1}),O(e_{2}))_{\tau}. \] Hence
$(\cdot,\cdot)_{\tau'}$ is again a generalized metric. We may use
the notation $(\cdot,\cdot)_{\tau'} = O (\cdot,\cdot)_{\tau}$.

\subsection{Factorizations of generalized metric, open-closed relations}
Let us start with a (different) generalized metric $\mathbf{H}$, described by a Riemannian metric $G$ and a $2$-form $\Phi$. Hence
\begin{equation} \label{genmetricH} \mathbf{H} = \bm{1}{\Phi}{0}{1} \bm{G}{0}{0}{G^{-1}} \bm{1}{0}{-\Phi}{0}. \end{equation}
Let $\theta$ be a $2$-vector field on $M$. The action of the $O(n,n)$ map $e^{-\theta}$ on the generalized metric $\mathbf{H}$ gives us a new generalized metric $\mathbf{G}$, which has the form
\begin{equation} \label{fact1} \mathbf{G} = \bm{1}{0}{\theta}{1} \bm{1}{\Phi}{0}{1} \bm{G}{0}{0}{G^{-1}} \bm{1}{0}{-\Phi}{1} \bm{1}{-\theta}{0}{1}. \end{equation}
By the previous discussion, there exists a unique Riemannian metric $g$ and a $2$-form $B$, such that
\begin{equation} \label{fact2} \mathbf{G} = \bm{1}{B}{0}{1} \bm{g}{0}{0}{g^{-1}} \bm{1}{0}{-B}{1}. \end{equation}
Comparing the two expressions (\ref{fact1}) and (\ref{fact2}) of
$\mathbf{G}$, we get the matrix equations
\begin{equation} \label{oc1} g - Bg^{-1}B = G - \Phi G^{-1} \Phi, \end{equation}
\begin{equation} \label{oc2} Bg^{-1} = \Phi G^{-1} - (G - \Phi G^{-1} \Phi) \theta, \end{equation}
which can be uniquely solved for $G$ and $\Phi$. Since $e^{-\theta}$ is invertible, we
can proceed the other way around as well. We also know how the
corresponding endomorphism $\tau_{\mathbf{H}}$ is changed by
$e^{-\theta}$. Namely, we have
\begin{equation} \tau_{\mathbf{G}} = e^{\theta} \tau_{\mathbf{H}} e^{-\theta}.\end{equation}
From that, we can easily find the relation between $+1$
eigenbundles:
\begin{equation} V_{+}^{\mathbf{G}} = e^{\theta} V_{+}^{\mathbf{H}}. \end{equation}
Since
\[ V_{+}^{\mathbf{G}} = \{ \xi + (g+B)^{-1}(\xi) \ | \ \xi \in T^{\ast}M \}, \]
and
\[ V_{+}^{\mathbf{H}} = \{ \xi + (G + \Phi)^{-1}(\xi) \ | \ \xi \in T^{\ast}M \}, \]
we get using the above formula that
\begin{equation} \label{formula_with_inverses}
(g+B)^{-1} = \theta + (G + \Phi)^{-1}. \end{equation} Formulae
(\ref{oc1}) and (\ref{oc2}) are the  symmetric and antisymmetric parts of (\ref{formula_with_inverses}).
 If $\theta$
is Poisson, (\ref{formula_with_inverses}) is the Seiberg-Witten
formula\footnote{For an earlier appearance of this type of formulae in the context of duality rotations see \cite{dufflu}.} relating closed and open string backgrounds in the
presence of a noncommutative structure represented by
$\theta$. In
particular, for given $g$, $B$ and $\theta$, we can find a unique
$G$ and $\Phi$, and conversely, for given $G$, $\Phi$ and $\theta$,
there exists a unique pair $g$ and $B$.

For $\Phi=0$ the open-closed relations can be given a slightly more geometric interpretation \cite{Asakawa:2012px}. Consider the inverse $\mathbf{G}^{-1}$ of the generalized metric $\mathbf{G}$. If we exchange the tangent and cotangent bundles $TM$ and $T^*M$, respectively, $\mathbf{G}^{-1}$ has the same properties as $\mathbf{G}$. Obviously, $\mathbf{G}^{-1}$ and $\mathbf{G}$ have identical graphs as well as $\pm 1$-eigenbundles. The open-closed relations, for $\Phi=0$, is a simple consequence of that.

\subsection{Dorfman bracket, Dirac structures, D-branes} \label{sec_dorfman}
Here we briefly recall some relevant facts concerning the Dorfman
bracket and Dirac structures, see, e.g., \cite{courant}, \cite{Gualtieri:2003dx}, \cite{Bouwknegt:2010zz}. Our vector bundle $E=TM \oplus T^{\ast}M$ can be equipped with
a structure of a Courant algebroid. The corresponding Courant
bracket is the antisymmetrization of the Dorfman bracket:
\begin{equation} \label{def_dorfman} [V+\xi,W+\eta]_{D} = [V,W] + \mathcal{L}_{V}(\eta) - \io_{W}(d\xi), \end{equation}
for all $(V+\xi) \in \Gamma(E)$. The corresponding pairing is the
canonical fiberwise metric (\ref{def_canmetric}).

A Dirac structure
is a (smooth) subbundle $L$ of $E$, which  is maximally isotropic with respect to
$\<\cdot,\cdot\>$ and involutive under the Dorfman bracket
(\ref{def_dorfman}).

Let $\theta$ be a rank-2 contravariant tensor field on $M$. As before,
define a vector bundle morphism $\theta: T^{\ast}M \rightarrow TM$ by
$\theta(\xi) = \theta(\cdot,\xi)$. Define a subbundle $G_{\theta}$ of $E$ as
its graph, that is
\begin{equation} G_{\theta} = \{ \xi + \theta(\xi) \ | \ \xi \in T^{\ast}M \}.
\end{equation}
It is known that $G_{\theta}$ is a Dirac structure with
respect to the Dorfman bracket, if and only if $\theta$ is a Poisson
bivector. Similarly, let $B$ be any rank-2 covariant tensor field on
$M$. Define $B(V) = B(V,\cdot)$ and its graph $G_{B}$ as
\begin{equation} G_{B} = \{ V + B(V) \ | \ V \in TM \}. \end{equation}
Again, one can show that $G_{B}$ is a Dirac structure, if and only if
$B$ is a closed $2$-form on $M$.

Furthermore, for any closed $B \in \Omega^{2}(M)$, one has
\begin{equation} \label{eB_dorfman} e^{B}[V+\xi,W+\eta]_{D} = [e^{B}(V+\xi), e^{B}(W+\eta)]_{D}, \end{equation}
and
\begin{equation} \label{eB_metric} \< e^{B}(V+\xi), e^{B}(W+\eta) \> = \< V+\xi, W+ \eta\>, \end{equation}
for all $(V+\xi),(W+\eta) \in \Gamma(E)$. In the other words, $e^{B}$ is an automorphism of the corresponding Courant algebroid. Note that (\ref{eB_dorfman}) is no longer true for $e^{\theta}$, where $\theta \in \Lambda^{2} \mathfrak{X}(M)$, but (\ref{eB_metric}) holds.

Generally, a Dirac structure $L$ provides a singular foliation of $M$ by presympletic leaves, which is generated by its image $\rho(L)$ of the Dirac structure under the anchor map.
We refer to \cite{Asakawa:2012px} for arguments in favor of the identification ``D-branes $\sim$ leaves of foliations defined by Dirac structures".  In the case we will consider later, $L$ will be given as a graph of a Poisson tensor $\theta$ and the corresponding foliation of $M$ will be the foliation generated by Hamiltonian vector fields, i.e., by symplectic leaves of $\theta$. Hence, in this case we will identify the symplectic leaves and D-branes.

\section{Gauge field as an orthogonal transformation of the generalized metric}
Let us start with a given Riemannian metric $g$ and $2$-form $B$.
Further, let $F$ be a $2$-form (at this point an arbitrary one\footnote{
Later, when discussing DBI action, $F$ will be closed and defined
only on a submanifold of $M$ supporting a D-brane. In which case,
all expression involving $F$ will make sense only when considered on
the D-brane.}). The gauge transformation defines new $2$-form $B' =
B + F$. To the pair $(g,B)$ corresponds the generalized metric
$\mathbf{G}$, see (\ref{fact2}). The generalized metric $\mathbf{G}'$ corresponding to
the pair $(g,B+F)$ has the following block matrix form:
\begin{equation}\label{fac1} \mathbf{G'} = \bm{1}{F}{0}{1}
\bm{1}{B}{0}{1} \bm{g}{0}{0}{g^{-1}} \bm{1}{0}{-B}{1}
\bm{1}{0}{-F}{1},\end{equation} that is, $\mathbf{G'}$ is related to
$\mathbf{G}$ by the $O(n,n)$ transform $e^{-F}$. As shown before, we
can always get $\mathbf{G}$ by action of $O(n,n)$ transformation
$e^{-\theta}$ on the generalized metric $\mathbf{H}$, where
$\mathbf{H}$ is described by fields $G$ and $\Phi$, see (\ref{genmetricH}).

One may ask, if there is a bivector $\theta'$ on $M$, such that we
get $\mathbf{G}'$ by the action of $e^{-\theta'}$ on the generalized
metric $\mathbf{H}'$, which is described by the same $G$ as
$\mathbf{H}$, but by gauged $2$-form $\Phi' = \Phi + F'$ for some
gauge field $F'$. This can be achieved under some assumptions,
however, only up to a certain additional $O(n,n)$ action. In
particular, there exists a vector bundle morphism $N: TM \rightarrow
TM$, such that

\begin{equation} \label{decompwithN} \mathbf{G}' = \bm{1}{0}{\theta'}{1} \bm{N^{T}}{0}{0}{N^{-1}}  \mathbf{H'} \bm{N}{0}{0}{N^{-T}} \bm{1}{-\theta'}{0}{1}, \end{equation}
where
\[ \mathbf{H'} = \bm{1}{\Phi'}{0}{1} \bm{G}{0}{0}{G^{-1}} \bm{1}{0}{-\Phi'}{1}. \]
Indeed, examine the block matrix decomposition:
\[ \mathbf{G}' = \bm{1}{F}{0}{1} \bm{1}{0}{\theta}{1} \bm{1}{\Phi}{0}{1} \bm{G}{0}{0}{G^{-1}} \bm{1}{0}{-\Phi}{1} \bm{1}{-\theta}{0}{1} \bm{1}{0}{-F}{1}.
\]
It suffices to consider the three rightmost matrices in the above
expression. Since we want to modify $\Phi$ to $\Phi + F'$, we may
proceed by inserting $1 = e^{-F'} e^{F'}$:
\[ \bm{1}{0}{-\Phi}{1} \bm{1}{-\theta}{0}{1} \bm{1}{0}{-F}{1} = \bm{1}{0}{-(\Phi + F')}{1} \bm{1}{0}{F'}{1} \bm{1}{-\theta}{0}{1} \bm{1}{0}{-F}{1}. \]
Now it is enough to note that the product of the last three
matrices, can be uniquely decomposed into a product of a diagonal
and an upper triangular block matrix---of course, only if we assume
that $(1 + \theta F)$ is invertible.  For this, use the
decomposition of $e^{-\theta} e^{-F}\in O(n,n)$ according to
(\ref{decomp}) as
\begin{equation} \label{exp_Ndecomposition} e^{-\theta} e^{-F} = e^{-F'} O_{N} e^{-\theta'}, \end{equation}
with $F' \in \Omega^{2}(M), \theta' \in \Lambda^{2} \mathfrak{X}(M)$
and $N \in \Gamma(\mbox{Aut}(TM))$. What we find are the following
expression for $\theta'$, $F'$ and $N$:
\begin{equation} \label{def_thetaprime}
\theta' = (1 + \theta F)^{-1}\theta = \theta(1 + F\theta)^{-1},
\end{equation}
 \begin{equation}\label{def_Fprime}
F' = F(1 + \theta F)^{-1} = (1 + F\theta)^{-1}F,
\end{equation}
\begin{equation} N = 1 + \theta F. \end{equation}
Comparing ({\ref{fac1}) and (\ref{decompwithN}), we get the
equalities
\begin{equation}\label{open-closedF1} g - (B+F)g^{-1}(B+F) =
N^{T} (G - (\Phi + F')G^{-1}(\Phi + F')) N \end{equation}and
\begin{equation} \label{oopen-closedF2}
(B+F)g^{-1} = N^T(\Phi+F') G^{-1}N^{-T} - N^T(G - (\Phi+F') G^{-1}
(\Phi+F'))N \theta' .
\end{equation}
Taking the determinant of (\ref{open-closedF1}), we find that
\begin{equation} \det( g - (B+F)g^{-1}(B+F) ) = \det(N)^{2} \cdot \det(G - (\Phi + F')G^{-1}(\Phi + F')). \end{equation}
This equality will play the central role when later discussing the
DBI action.

Furthermore, following the same type of arguments leading to
(\ref{formula_with_inverses}) we see that the equations
(\ref{open-closedF1}) and (\ref{oopen-closedF2}) can equivalently be
written as
\begin{equation} \label{formula_with_inversesF}
(g+B+F)^{-1} = \theta' + (N^T(G + \Phi+F')N)^{-1}. \end{equation}

Finally, let us examine the objects $F'$ and $\theta'$ using the
tools described in subsection \ref{sec_dorfman}. We will concentrate
on the case important for the discussion of the DBI action and
noncommutative gauge theory. Therefore, in the rest of this section,
we assume that $\theta$ is Poisson and $F$ is closed. $\theta'$ is a bivector on $M$. 
For the graphs of $\theta$ and $\theta'$ we have
\begin{equation} e^{F} G_{\theta} = G_{\theta'}. \end{equation}
Since $e^{F}$ is an automorphism of Dorfman bracket, $G_{\theta'}$
has to be again a Dirac structure of $E$. Hence, $\theta'$ is a
Poisson bivector. Similarly, one can see that
\begin{equation} e^{\theta} G_{F} = G_{F'}. \end{equation}
This is no more an automorphism of Dorfman bracket but it preserves
the (maximal) isotropy property of $G_{F}$. Hence $G_{F'}$ is an
isotropic subbundle of $E$ and $F'$ is therefore a $2$-form on $M$.
Let us also note, that $F'$ doesn't need to be closed. The last
remark: In case that $(1 + \theta F)$ is not invertible, $e^{F}
G_{\theta}$ still makes perfect sense as a Dirac structure.
Similarly, $e^{\theta} G_{F}$ will still define an almost Dirac
structure.

\section{Non-topological Poisson-sigma model and Polyakov action}
In this section we review the non-topological Poisson-sigma model
from the generalized geometry point of view developed in the
previous sections.

Let us consider a $2$-dimensional world-sheet $\Sigma$ with a set
of local coordinates $(\sigma^{0},\sigma^{1})$. We assume that
$\sigma^{\mu}$ are Cartesian coordinates for a Lorentzian metric $h$
with signature $(-,+)$ on $\Sigma$. Furthermore, we consider an
$n$-dimensional target manifold $M$, equipped with a metric $G$,
$2$-vector $\theta$ and a $2$-form $\Phi$. We can assume $\Sigma$ with a non-empty boundary $\partial \Sigma$. On $M$ assume an abelian
gauge field $A$ coupling to the boundary (and extending to $\Sigma$,
the field strength being $F=dA$). We also choose some local
coordinates $(y^1,\dots, y^n)$ on $M$. Lower case Latin characters
will always correspond to these coordinates. For the components of
the smooth map $X: \Sigma \rightarrow M$ we will use the following
notation: $X^{i} = y^{i}(X)$. In this section it will be convenient
to introduce the following notation: We put $\bar G :=N^TGN$, $\bar
\Phi :=N^T\Phi N$ and $\bar F' :=N^TF'N$ and introduce auxiliary
fields $\eta_{i}$ and $\~\eta_{j}$, which transform under change of
local coordinates on $M$ according to their index structure. We
combine them in a $2n$-dimensional column vector $\Psi^T :=(\eta, \tilde
\eta)$. We also introduce another $2n$-dimensional column vector $V^T:=
(\partial_0 X,
\partial_1 X)$. Finally, we define a $2n \times 2n$ matrix\footnote{Here, we neither need to assume that $\theta$ is Poisson nor that $F$ is closed.}
\begin{equation}
\bar{\mathbf{G}}=\begin{pmatrix} -\bar G &  -\bar\Phi - \bar F' \\
\bar \Phi +\bar F'& \bar G\end{pmatrix}^{-1} +
\begin{pmatrix} 0 & \theta' \\
-\theta' & 0\end{pmatrix}. \end{equation}
Our (non-topological)
Poisson-sigma model action is
\begin{equation} \label{def_action}
S[\eta,\~\eta,X] := \int d^{2}\sigma \frac{1}{2} \Psi^T
\bar{\mathbf{G}}\Psi + \Psi^TV.
\end{equation}
Using relations (\ref{open-closedF1}), (\ref{oopen-closedF2}), the
action (\ref{def_action}) can equivalently be written as
\begin{equation} \label{def_action1}
S[\eta,\~\eta,X] := \int d^{2}\sigma \frac{1}{2} \Psi^T
\tilde{\mathbf{G}}\Psi + \Psi^TV,
\end{equation}
where
\begin{equation}
\tilde{\mathbf{G}}=\begin{pmatrix} -g & -B-F \\
B+F& g\end{pmatrix}^{-1}\end{equation} with $g$, $B$ and $F$
being related to $G$, $\Phi$ and $F'$ by (\ref{open-closedF1}),
(\ref{oopen-closedF2}) and (\ref{def_Fprime}). Integrating out the
auxiliary fields $\eta$ and $\tilde \eta$ we obtain the Polyakov
action expressed equivalently either in open or closed variables
\begin{equation} \label{Polyakov}
S[X] := - \frac{1}{2}\int d^{2}\sigma  V^T \bar{\mathbf{G}}^{-1}V =
- \frac{1}{2}\int d^{2}\sigma  V^T \tilde{\mathbf{G}}^{-1}V.
\end{equation}
Actually, all this can be seen rather straightforwardly. For this,
note that relations (\ref{open-closedF1}), (\ref{oopen-closedF2})
can alternatively be expressed as the equality of matrices
$\bar{\mathbf{G}}= \tilde{\mathbf{G}}$. The relations in the form
(\ref{formula_with_inversesF}) and their transposes are obtained
from the nonzero off-diagonal blocks
after the similarity transformation with the block matrix $\begin{pmatrix} 1 & 1\\
1& -1\end{pmatrix}$ is applied to the equality $\bar{\mathbf{G}}=
\tilde{\mathbf{G}}$.

The generalized metric $\mathbf{G}'$ can explicitly be seen either
in the Hamiltonian corresponding to the Polyakov action
(\ref{Polyakov}) or in the Hamiltonian corresponding to the action
(\ref{def_action}) after the equations of motions for one half of
the auxiliary fields, the $\tilde \eta$s, are used. As can
straightforwardly be checked, these Hamiltonians are identical. To
write down the result we introduce a new $2n$-dimensional column
vector $\Upsilon^T:=(
\partial_1 X,\eta)$. The auxiliary fields $\eta_i$ become the canonical
momenta and the Hamiltonian is
\begin{equation}\label{Ham}
H[X,\eta]=\frac{1}{2} \int d\sigma^1 \Upsilon^T \mathbf{G}' \Upsilon\,,
\end{equation}
where $\mathbf{G}'$ the matrix given by the two equivalent
decompositions (\ref{fac1}) and (\ref{decompwithN}). Hence, we have
the same Hamiltonian using either the closed or the open variables.
Let us note that, for $F=0$, the relation between the action  (\ref{def_action}) and the action
(\ref{Polyakov}) with $G'$ expressed as in (\ref{fac1}) can be found in \cite{Baulieu:2001fi}. The Hamiltonian (\ref{Ham}) with $G'$ given by
(\ref{decompwithN}) can be found, again for $F=0$, in \cite{halmagyi2}. Polyakov actions like the first one in (\ref{Polyakov}) appeared (with $F=0$) in \cite{Klimcik:1995dy} in the context of Poisson-Lie T-duality.

\section{Seiberg-Witten map}
For an approach to the non-abelian case, using cohomological methods akin to the ones of Zumino's famous decent equations \cite{Zumino:1984ws}, see \cite{cohoswmap,swnoncomm}. Here we follow the approach of \cite{Jurco:2000fb}, \cite{Jurco:2000fs}, \cite{Jurco:2001my}, where it was shown that the Seiberg-Witten field redefinition from the commutative to the non-commutative setting has its origin in a change of coordinates given by a map $\rho: M \rightarrow M$, such that $\rho^{\ast}(\theta') = \theta$.\footnote{As said before, here we assume  only topologically trivial $[H]$-flux. The interested reader may find some relevant discussion concerning nontrivial $H$ and the related non-commutative gerbe in \cite{NCgerbe}.} This map can be derived using a  generalization of Moser's lemma:
Consider the family of Poisson bivectors
\begin{equation} \label{thetat}
\theta_{t} = \theta (1 + t F \theta)^{-1} \end{equation} parameterized by $t
\in [0,1]$. Of course, we have to presume that the formula is well-defined. To see that these $\theta_t$ are indeed Poisson for all $t$, simply
observe that $G_{\theta_{t}} = e^{tF} G_{\theta}$ holds for the respective graphs.\footnote{Let us
note again that $e^{tF} G_{\theta}$ is a bona-fide Dirac structure even
for non-invertible $(1+tF\theta)$.}
Partial differentiation of (\ref{thetat}) with respect to $t$ leads
to the differential equation
\[ \partial_{t} \theta_{t} = -\theta_{t} F \theta_{t}. \]
For $F = dA$, this can be rewritten as
\[ \partial_{t} \theta_{t} = - \mathcal{L}_{\theta_{t}(A)} \theta_{t}, \]
with a vector field $\theta_{t}(A) := \theta_{t}(\cdot,A)$, with initial condition
$\theta_{0} = \theta$. This differential equation can be integrated
to a flow $\phi_{t}$, such that $\phi_{t}^{\ast}(\theta_{t}) =
\theta$. Thus $\rho = \phi_{1}$. Obviously, $\rho$ explicitly
depends on the choice of  gauge potential $A$, hence we shall use the notation
$\rho_{A}$. To avoid possible
confusion, we will for a moment notationally distinguish between the
tensor itself and its components in coordinates. Therefore we
introduce the matrix $(\bs{\theta})^{ij} := \theta^{ij}$. Also,
denote ${J^{i}}_{k} = \frac{\partial \rho^{i}}{\partial x^{k}}$. We
have
\[ \rho_A^{\ast}( {\theta'}^{kl} ) = {J^{k}}_{i} {J^{l}}_{j} \theta^{ij}. \]
We thus get that
\begin{equation} \label{eq_determinanty}
\det{ \rho_A^{\ast}(\bs{\theta'})} = J^{2} \det{ \bs{\theta}}.
\end{equation}
Let us assume for a moment that $\bs{\theta}$ is invertible. From
(\ref{def_thetaprime}) we see that so is $\rho^{\ast}_A
\bs{\theta'}$. We immediately have that
\begin{equation}\label{Jac} J^{-2} = \det{( \bs{\theta}
(\rho^{\ast}_A\bs{\theta'})^{-1} )}. \end{equation} For degenerate
$\theta$ and hence also $\theta'$ the formula (\ref{Jac}) still
makes sense and we can  argue as follows: Since the map
$\rho_{A}$ is infinitesimally generated by the vector field
$\theta_{t}(A)$, and the kernels of all $\theta_{t}$'s are the same, we
see that $\rho_{A}$ only changes coordinates on the symplectic
leaves (of $\theta$). We can thus restrict ourselves to the non-degenerate case in
order to carry out the computation of the Jacobian.

In the next section, we will discuss the case when the Poisson structure $\theta$ (i.e., the corresponding Dirac structure) will be used, following the suggestion of \cite{Asakawa:2012px}, to define the D-branes as its symplectic leaves. The above argument shows that we can safely restrict our discussion without the loss of generality to any of the respective D-branes. In such a case, the (Seiberg-Witten) map $\rho_A$  is a diffeomorphism of the D-brane world-volume $D$. The Poisson structures $\theta_t$ have in fact the same symplectic foliations for all $t$. Actually,  all Poisson structures $\theta_t$, including in particular $\theta$ and
$\theta'$, are  Morita equivalent \cite{2002math......2099B}.

Finally, on the level of Dirac structures, the Seiberg Witten map is the map of graphs $\rho^*: G_\theta' \mapsto G_\theta$. More explicitly,
$$\{\theta' (\eta) +\eta ,\eta \in T^* M\} \mapsto
\{\theta (\eta) + \eta , \eta \in T^* M\}=\{ N\theta' N^{-T}(\eta) + N^{-T}(\eta), \eta \in T^* M\}.$$
Hence,  the Seiberg-Witten map can be seen as the map induced by
the $O(n,n)$ transformation $O_N$ entering the decomposition (\ref{decomp}), if one considers D-branes which are symplectic leaves.

\section{Noncommutative gauge theory and DBI action}
In the previous sections we have described all ingredients needed
for our discussion of noncommutativity of D-branes as a consequence
of their generalized geometry. Namely, we have seen that the
relations (\ref{oc1}), (\ref{oc2}), (\ref{open-closedF1}) and the
(semiclassical) Seiberg-Witten have their root in generalized
geometry. Actually, it is know for quite some time \cite{Jurco:2001my} that the
equivalence of the commutative and (semiclassically) noncommutative
DBI actions follows once one has established (\ref{oc1}),
(\ref{oc2}), (\ref{open-closedF1}) and has understood the
(semiclassical) Seiberg-Witten map as a (local) D-brane
diffeomorphism. Nevertheless, according to our best knowledge, the direct
relation to generalized geometry is new. Moreover, the discussion generalizes to
the case of $M$-theory branes \cite{Jurco:2012yv}, \cite{Schupp:2012nq} and will be elaborated in detail in a forthcoming paper.
Here we will include the derivation of the equivalence of the
commutative and (semiclassically) noncommutative DBI actions for the
sake of completeness and the reader's convenience. For related work based on dualities, see \cite{dualityinvariant}. 

Assume that we have a D-brane $D$ of dimension $d$, i.e, a submanifold of target
space-time $M$ equipped with a line bundle with a connection $A$ and
corresponding  field strength $F$. Also, consider the
restrictions (pullbacks) of the background fields (open and closed
ones) to $D$. While describing the Seiberg-Witten map in the previous section, we have seen that it is quite natural to assume that there is a relation between the D-brane and the Poisson tensor $\theta$.\footnote{Recall, in accordance with our above discussion of the open-closed relations, here we start from a given closed background $(g, B)$, pick a $\theta$ and determine uniquely the open variables $(G,\Phi)$.} Namely, assume that our D-brane is of a particular kind, i.e., one which comes as symplectic leaf of the Poisson structure $\theta$.\footnote{It is straight-forward to modify everything to the case where the D-brane is a submanifold, such that the restriction of $\theta$ to it defines a regular Poisson structure, i.e. a Poisson structure having constant rank.}
As argued before, under this assumption, the Seiberg-Witten map is a D-brane diffeomorphism.

Before we turn to the discussion of the DBI action and its
commutative and noncommutative description, we discuss the
relation between the effective closed and open string coupling
constants $g_s$ and $G_s$, respectively \cite{Seiberg:1999vs}. These are related as
\[G_s = g_s \Big(\frac{\det (G+\Phi)}{\det (g + B)}\Big)^{1/2}.\]
We can use the formula for the
determinant of a sum of a symmetric matrix $S$ and an antisymmetric
matrix $A$, $|S+A|=|S|^{1/2}|S-AS^{-1}A|^{1/2}$, and the relation
(\ref{oc1}) to rewrite this as
\begin{equation}\label{coupling}
G_s = g_s \Big(\frac{\det G}{\det g}\Big)^{1/4}.
\end{equation}
A most intriguing relation is
obtained from (\ref{coupling}) and the relation
(\ref{open-closedF1}), again using the above mentioned formula for
the determinant of a sum of a symmetric and an antisymmetric matrix:
\begin{equation}\label{DBI1}
\frac{1}{g_s} \det\,\hskip-0.15cm^{1/2}(g+B+F) = \frac{1}{G_s}
\det\,\hskip-0.15cm^{1/2}(1+\theta F)
\det\,\hskip-0.15cm^{1/2}(G+\Phi + F').\end{equation} Integrating
over the D-brane world-volume
\begin{equation}
\int d^dx \frac{1}{g_s} \det\,\hskip-0.15cm^{1/2}(g+B+F) = \int d^dx
\frac{1}{G_s} \det\,\hskip-0.15cm^{1/2}(1+\theta F)
\det\,\hskip-0.15cm^{1/2}(G+\Phi + F'),
\end{equation}
recalling (\ref{Jac}), and performing the change of coordinates
according to the Seiberg-Witten map, we finally obtain a
relation between the commutative and semiclassically noncommutative
DBI actions
\begin{equation}\label{DBI2}
S^c_{DBI}:=\int d^dx \frac{1}{g_s} \det\,\hskip-0.15cm^{1/2}(g+B+F)
= \int d^dx \frac{1}{\hat G_s}
\det\,\hskip-0.15cm^{1/2}\Big(\frac{\hat \theta}{\theta}\Big)
\det\,\hskip-0.15cm^{1/2}(\hat G+\hat\Phi + \hat F')=:S^{nc}_{DBI}.
\end{equation}
The hat ``$\hat{\mbox{ }}$'' has the following meaning: On matrix
elements of $\theta$ it is defined as $\hat \theta^{ij}:=\rho^*_A
(\theta^{ij})$, and similarly for the other objects. As a result of this definition, $\hat F'$
is the semiclassically noncommutative field strength, which under
the gauge transformation $\delta A = d\lambda$ transforms
semiclassically noncommutatively, i.e.,
$$\delta \hat
F'_{ij} = \{\hat F'_{ij},\tilde\lambda\},$$
$$\tilde\lambda=\sum
\frac{(\theta_t(A)+ \partial_t)^n (\lambda)}{(n+1)!}|_{t=0}.$$ Here,
the curly bracket is the Poisson bracket corresponding to the
Poisson tensor $\theta$ and $\tilde \lambda$ is the (semiclassical)
noncommutative gauge parameter.

Let us note: The commutative DBI action $S^c_{DBI}$ on the LHS in
(\ref{DBI2}) is the effective D-brane action obtained from the
Polyakov action (\ref{Polyakov}). Expressed directly in terms of the
matrix $\tilde{\mathbf{G}}$, the action $S^c_{DBI}$ is the integral of
\begin{equation}
\det\,\hskip-0.15cm^{1/4}
\tilde{\mathbf{G}}
\end{equation}
up to the inverse of the closed coupling constant ${g_s}$.
Hence, an alternative---but completely equivalent---way of obtaining
the relation between the commutative and semiclassically
noncommutative DBI actions (\ref{DBI2}) is to start from the matrix
equality $\tilde{\mathbf{G}}=\bar{\mathbf{G}}$. This makes the
relation to the Polyakov action more transparent. We leave the
details to the reader.

Finally, the Hamiltonian (\ref{Ham}) can equivalently be expressed
using either the ``commutative" (\ref{fac1}) or ``noncommutative''
(\ref{decompwithN}) decompositions
 of the generalized metric $\mathbf{G}'$. This is maybe the most
 direct hint from generalized geometry about the necessity of a
 relation like (\ref{DBI2}).

\section*{Acknowledgement}
We would like to dedicate this article to Bruno Zumino on the occasion of his 90th birthday. We would like to thank Satoshi Watamura for important comments on
an earlier version of the manuscript.
The research of B.J. was supported by grant GA\v CR P201/12/G028.
The research of J.V. was supported by Grant Agency of the Czech
Technical University in Prague, grant No. SGS13/217/OHK4/3T/14. We
also thank to DAAD (PPP) and ASCR \& MEYS (Mobility) for supporting our collaboration and
gratefully acknowledge support from the DFG within the Research Training Group 1620 ``Models of Gravity''.

\bibliography{nambu_sigma}
\end{document}